\documentclass[fleqn,12pt,twoside]{article}

\usepackage{espcrc1}
\usepackage{graphicx}
\usepackage[figuresright]{rotating}

\newcommand{\AmS}{{\protect\the\textfont2
  A\kern-.1667em\lower.5ex\hbox{M}\kern-.125emS}}
\title{Physics potential of beta-beams}

\author{C. Volpe \address{Institut de Physique Nucl\'eaire, \\ 
        F-91406 Orsay cedex, France}}
       
\begin{document}

\maketitle

\begin{abstract}
Beta-beams is a novel concept for the production of neutrino beams exploiting
boosted radioactive ions which decay through beta-decay. Here we describe a 
project, 
currently under investigation, where 
such beams are used to address the issue of CP violation in the 
lepton sector. We also mention the interest of having low-energy beta-beams.
The connection to projects for the production of very intense exotic ion beams,
like EURISOL, is emphasized.

\end{abstract}

\section{Introduction}

Nuclei represent a precious laboratory for the study of fundamental
issues like neutrino properties. Pauli first postulated
the existence of these elusive particles to explain the missing energy
observed in nuclear beta decay. 
A very timely example is offered by
the present search for the absolute value of the neutrino mass
through tritium beta-decay.
The important issue of the Dirac or Majorana nature of neutrinos
is addressed by experiments that look for neutrinoless double-beta decay
in nuclei \cite{review}.

Neutrino oscillation experiments have shown nowadays that neutrinos are massive
particles. This phenomenon depends on the so-called neutrino mixing angles
and on the difference between the square of the masses.
These experimental observations are accounted for by introducing
the Maki-Nakagawa-Sakata-Pontecorvo (MNSP) matrix which relates
the neutrino flavor  to the mass basis, like the Cabibbo-Kobayashi-Maskawa (CKM) matrix in the quark sector.  This unitary matrix depends on
the three mixing angles and one (Dirac) phase introducing CP violation in
the lepton sector.
In order to measure CP violating effects, one needs to perform comparative
studies of neutrino versus anti-neutrino oscillations like, e.g.
comparing $\nu_e \rightarrow \nu_{\mu}$
with $\bar{\nu}_e \rightarrow \bar{\nu}_{\mu}$. 
Several other important issues are still open on neutrino properties, 
among which their Dirac versus Majorana nature, the
electromagnetic properties or
the absolute mass scale \cite{review}.

Various methods to produce intense neutrino beams are now explored, based on
 intense 
 conventional (from the decay of pions and muons) beams - 
 also called super-beams - 
or on new concepts, like neutrino factories \cite{review}.
In this respect, 
nuclei can offer a precious tool 
through the novel concept of beta-beams:
 pure (anti-)electron neutrino beams produced by accelerating
radioactive ions that decay through beta-decay \cite{zucchelli}.

\section{The Beta-beam Project}

Beta-beams have been the starting point for a new project at CERN, at present under
investigation.
According to the first feasibility study \cite{zucchelli,mats},
the neutrino beams are obtained by producing, collecting and accelerating
the ions, first at several hundred MeV and then
to GeV energies, by injection in the (already existing)
PS and SPS.
At present, the best candidates
as   $\bar{\nu}_e$ (${\nu}_e$)  emitters
are $^{6}$He ($^{18}$Ne).
The (anti-)neutrino energies are tuned to meet the neutrino oscillation
frequencies
(note that the ion Lorentz gamma factor is related to the neutrino energy
according to $E_{\nu}=2\gamma Q_{\beta}$ with $Q_{\beta}$ being the beta
Q-value).
Once the ions have reached the required energies,
they are injected in a storage ring with long straight sections (2.5 km
for a total length of 7 km). Such a ring
needs to be built \cite{zucchelli,mats}.
All the ions 
(about $2.~10^{14}$ $^{6}$He and $10^{13}$ $^{18}$Ne) 
along the straight sections produce neutrino beams
that are fired to far detectors for the comparative study of
$\nu_e \rightarrow \nu_{\mu}$
and $\bar{\nu}_e \rightarrow \bar{\nu}_{\mu}$ oscillations.
The production and first acceleration steps present
strong links to the european EURISOL conceptual design. This has resulted
in a common bid for a design study withing the 6th European Union Framework
program.

Various scenarios to perform CP violation studies are being proposed
\cite{zucchelli,mats,mauro,jj,tmms,migliozzi}.
In the first scenario \cite{zucchelli,mats,mauro,migliozzi},
beta-beams
 are fired to a
Cherenkov detector like UNO \cite{uno}
(440 ktons fiducial volume,  20 times bigger than the
present Super-Kamiokande detector), located in an (upgraded)
Fr\'ejus Underground Laboratory, 130 km from CERN.
To match this distance, the ions have to reach  
a Lorentz factor $\gamma$=100 (60) for  $^{18}$Ne ($^{6}$He).
Note that 
the gigantic detector can be used 
to improve the present sensitivity on proton decay and
as a core-collapse supernova neutrino observatory.
Other interesting
scenarios have been proposed where the ions are accelerated to
much higher energies and sent to further distancies, like e.g. to the Gran
Sasso laboratory \cite{jj,tmms}.
The feasibility of the very high-gamma scenarios
needs further studies.
Figure 1 shows the CP sensitivity
that can be achieved in the different scenarios,
as a function of the neutrino mixing angle $\theta_{13}$ \cite{jj}.
Finally, this project has the attractive feature that
if conventional beta-beams (producing both $\nu_{\mu}$ and
$\bar{\nu}_{\mu}$)
are sent to the same detector, T and CPT violation studies
can also be addressed by comparing
$\nu_e \rightarrow \nu_{\mu}$ with
$\nu_{\mu} \rightarrow \nu_{e}$
and $\bar{\nu}_{\mu} \rightarrow \bar{\nu}_{e}$ 
oscillations respectively
\cite{mauro}.

\begin{figure}[h]
\begin{center}
\includegraphics[angle=0.,scale=0.4]{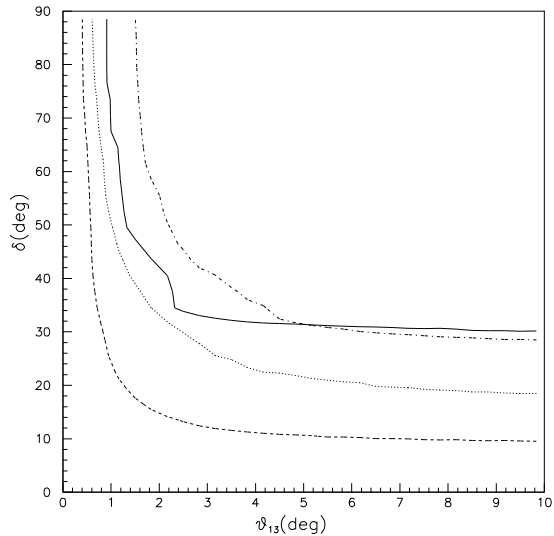}
\end{center}
\caption{CP sensitivity -defined as the ability to discriminate
between $\delta=0^{\circ}$ and $180^{\circ}$ with 99$\%$ C.L.-
for the following different scenarios: CERN to Fr\'ejus,
with $\gamma$=100 (60) for  $^{18}$Ne ($^{6}$He)
and a UNO-type detector (solid line);
CERN to Gran Sasso, with $\gamma$=580 (350) for  $^{18}$Ne ($^{6}$He)
and UNO  (dashed line) or a factor 10 smaller (dashed-dotted
line);
CERN to Canary islands, with $\gamma$=2500 (1500) for  $^{18}$Ne ($^{6}$He)
and a 40 kton detector \cite{jj}.}
\label{fig:jj}
\end{figure}

\section{Low-energy Beta-beams}

The beta-beam concept can also be used to
establish a facility producing neutrinos 
with low energies (i.e. several tens
of MeV up to a hundred) \cite{lownu}.
Such a facility can be either part of the high
energy facility at CERN,
or part of one of the laboratories producing
intense radioactive ion beams in the future \cite{lownu}.
The option of a low-energy neutrino facility based on conventional
beams is also at present under study \cite{orland}.
In \cite{gailnew}
a detailed comparison between the two options for the study
of neutrino interactions on lead 
is presented.

The availability of low energy neutrino beams would offer the opportunity
to address
a rich physics program \cite{lownu,orland,jpg}.
In \cite{munu} the potential of low-energy beta-beams is explored
as far as the neutrino magnetic moment is concerned.
In this case the ions are produced, collected
and used as an intense neutrino source at rest, inside a
$4 \pi$ detector.
It is shown that one might improve present upper limits, obtained
with direct measurements,
 by almost an order of magnitude.
 Sistematic studies on
neutrino-nucleus interactions
can also be performed e.g. by putting
a close detector to the storage ring where low-energy
beta-beams are collected \cite{julien}.
This is a topic of current great interest for various
domains of physics \cite{jpg,revuekubo,nu2004}.
Table I shows neutrino-nucleus interaction rates expected
for $\gamma$=14 and 
a small or a large storage rings.
The former
corresponds to the one planned for the future GSI
facility while the latter is the ring included in 
the beta-beam baseline scenario at CERN (Details can be found in \cite{julien}.).
An advantage of beta-beams is that one can vary the average neutrino
energy by accelerating the ions to different gammas.
From Table I one can see that interesting neutrino-nucleus
interaction rates can be otained 
by using typical parameters available from
existing feasibility studies. Therefore,
low-energy beta-beams could offer a unique opportunity
to perform neutrino-nucleus studies of interest for nuclear
physics, particle physics and astrophysics \cite{lownu,julien}.

In conclusion, these projects are very promising. 
A detailed feasibility study 
will be performed,
jointly with the EURISOL design study, 
in the coming years.

\begin{table*}
\caption{}
\label{table:1}
\newcommand{\m}{\hphantom{$-$}}
\newcommand{\cc}[1]{\multicolumn{1}{c}{#1}}
\renewcommand{\tabcolsep}{2pc} 
\renewcommand{\arraystretch}{1.2} 
\begin{tabular}{@{}lllll}
\hline
 Reaction          &  Ref.            & Mass  & Small Ring & Large Ring \\
                      & & (tons) & ($L$=450 m, $D$= 150 m) & ($L$=7 km, $D$=2.5 km) \\   
  \hline
  $\nu +$D          &\cite{revuekubo}  &  \m35   &  \m2363      & \m180
  \\
   $\bar\nu +$D      &\cite{revuekubo}  &  \m35   &  \m25779     &
    \m1956      \\
      $\nu + ^{16}$O    &\cite{nuO}        &  \m952  &  \m6054      &
        \m734       \\
	   $\bar\nu +^{16}$O &\cite{nuO}        &  \m952  &  \m82645     &
	      \m9453      \\
	      $\nu +^{56}$Fe    &\cite{nuFe}       &  \m250  &  \m20768 &
	          \m1611      \\
		       $\nu +^{208}$Pb   &\cite{volpelead}       &  \m360  &  \m103707
		            &  \m7922      \\
			    \hline
			    \end{tabular}\\[2pt]
Neutrino rates (events/year)
on deuteron, oxygen, iron and lead \cite{julien},
with $\gamma=14$
for the parent ion. The cross sections are taken
from referred references.
The detectors are located at 10 meters from the storage ring and have cylindrical
shapes ($R$=1.5 m and $h$=4.5 m for deuteron, iron and lead,
$R$=4.5 m and $h$= 15 m for oxygen, where $R$ is the radius and $h$ is the
depth of the detector).
Their mass is indicated in the second column. Rates obtained for two 
different
storage ring sizes are presented ($L$ is the total
length and $D$ is the length of the straight sections). Here 1 year = $3.2
\times 10^{7}$ s.  
\end{table*}

\end{document}